
%
%
%
%
%
%
%
%
\def\standardrisposta{s }\def\reducedrisposta{r }
\def\mplarisposta{mpla }
\def\doublerisposta{d }\def\cartarisposta{e }\def\amsrisposta{y }
\newcount\ingrandimento \newcount\sinnota \newcount\dimnota
\newcount\unoduecol \newdimen\collhsize \newdimen\tothsize
\newdimen\fullhsize \newcount\controllorisposta \sinnota=1
\newskip\infralinea  \global\controllorisposta=0
\message{ ********    Welcome to PANDA macros (Plain TeX, AP, 1991)}
\message{ ******** }
\message{       You'll have to answer a few questions in lowercase.}
\message{>  Do you want it in double-page (d), reduced (r)}
\message{or standard format (s) ? }\read-1 to\risposta
\message{>  Do you want it in USA A4 (u) or EUROPEAN A4 (e)}
\message{paper size ? }\read-1 to\srisposta
\message{>  Do you have AMSFonts 2.0 (math) fonts (y/n) ? }
\read-1 to\arisposta
%
%
%
%
\ifx\risposta\standardrisposta \ingrandimento=1200
\message{>> This will come out UNREDUCED << }
\dimnota=2 \unoduecol=1 \global\controllorisposta=1 \fi
\ifx\risposta\reducedrisposta \ingrandimento=1095 \dimnota=1
\unoduecol=1  \global\controllorisposta=1
\message{>> This will come out REDUCED << } \fi
\ifx\risposta\doublerisposta \ingrandimento=1000 \dimnota=2
\unoduecol=2  \global\controllorisposta=1 
\message{>> You must print this in LANDSCAPE orientation << } \fi
\ifx\risposta\mplarisposta \ingrandimento=1000 \dimnota=1
\message{>> Mod. Phys. Lett. A format << }
\unoduecol=1 \global\controllorisposta=1 \fi
\ifnum\controllorisposta=0  \ingrandimento=1200
\message{>>> ERROR IN INPUT, I ASSUME STANDARD UNREDUCED FORMAT <<< }
\dimnota=2 \unoduecol=1 \fi
\magnification=\ingrandimento
%
%
%
%
\newdimen\eucolumnsize \newdimen\eudoublehsize \newdimen\eudoublevsize
\newdimen\uscolumnsize \newdimen\usdoublehsize \newdimen\usdoublevsize
\newdimen\eusinglehsize \newdimen\eusinglevsize \newdimen\ussinglehsize
\newskip\standardbaselineskip \newdimen\ussinglevsize
\newskip\reducedbaselineskip \newskip\doublebaselineskip
\eucolumnsize=12.0truecm    
\eudoublehsize=25.5truecm   
\eudoublevsize=6.5truein    
\uscolumnsize=4.4truein     
\usdoublehsize=9.4truein    
\usdoublevsize=6.8truein    
\eusinglehsize=6.5truein    
\eusinglevsize=24truecm     
\ussinglehsize=6.5truein    
\ussinglevsize=8.9truein    
\standardbaselineskip=16pt  
\reducedbaselineskip=14pt   
\doublebaselineskip=12pt    
%
%
\def\Portoffset{}
\def\Landoffset{}
\ifx\risposta\mplarisposta \def\Portoffset{\hoffset=1.8truecm} \fi
%
%
%
\tolerance=10000
\parskip 0pt plus 2pt  \leftskip=0pt \rightskip=0pt
%
%
\ifx\risposta\standardrisposta \infralinea=\standardbaselineskip \fi
\ifx\risposta\reducedrisposta  \infralinea=\reducedbaselineskip \fi
\ifx\risposta\doublerisposta   \infralinea=\doublebaselineskip \fi
\ifx\risposta\mplarisposta     \infralinea=13pt \fi
\ifnum\controllorisposta=0    \infralinea=\standardbaselineskip \fi
\ifx\risposta\doublerisposta   \Landoffset \else \Portoffset \fi
\ifx\risposta\doublerisposta \ifx\srisposta\cartarisposta
\tothsize=\eudoublehsize \collhsize=\eucolumnsize
\vsize=\eudoublevsize  \else  \tothsize=\usdoublehsize
\collhsize=\uscolumnsize \vsize=\usdoublevsize \fi \else
\ifx\srisposta\cartarisposta \tothsize=\eusinglehsize
\vsize=\eusinglevsize \else  \tothsize=\ussinglehsize
\vsize=\ussinglevsize \fi \collhsize=4.4truein \fi
\ifx\risposta\mplarisposta \tothsize=5.0truein
\vsize=7.8truein \collhsize=4.4truein \fi
%
%
%
%
\newcount\contaeuler \newcount\contacyrill \newcount\contaams
\font\ninerm=cmr9  \font\eightrm=cmr8  \font\sixrm=cmr6
\font\ninei=cmmi9  \font\eighti=cmmi8  \font\sixi=cmmi6
\font\ninesy=cmsy9  \font\eightsy=cmsy8  \font\sixsy=cmsy6
\font\ninebf=cmbx9  \font\eightbf=cmbx8  \font\sixbf=cmbx6
\font\ninett=cmtt9  \font\eighttt=cmtt8  \font\nineit=cmti9
\font\eightit=cmti8 \font\ninesl=cmsl9  \font\eightsl=cmsl8
\skewchar\ninei='177 \skewchar\eighti='177 \skewchar\sixi='177
\skewchar\ninesy='60 \skewchar\eightsy='60 \skewchar\sixsy='60
\hyphenchar\ninett=-1 \hyphenchar\eighttt=-1 \hyphenchar\tentt=-1
%
\font\tencmmib=cmmib10  \newfam\cmmibfam  \skewchar\tencmmib='177
\font\tencmbsy=cmbsy10  \newfam\cmbsyfam  \skewchar\tencmbsy='60
\def\scaps{\cmcsc}                 
\font\tencmcsc=cmcsc10  \newfam\cmcscfam
\ifnum\ingrandimento=1095

\font\capsone=cmcsc10 at 10.95pt 

\else

\font\capsone=cmcsc10 at 12pt 
\fi

\def\ttaarr{\bf}		
\def\ppaarr{\sl}		

%
%
\newfam\eufmfam \newfam\msamfam \newfam\msbmfam \newfam\eufbfam
\def\Loadeulerfonts{\global\contaeuler=1 \ifx\arisposta\amsrisposta
\font\teneufm=eufm10              
\font\eighteufm=eufm8 \font\nineeufm=eufm9 \font\sixeufm=eufm6
\font\seveneufm=eufm7  \font\fiveeufm=eufm5
\font\teneufb=eufb10              
\font\eighteufb=eufb8 \font\nineeufb=eufb9 \font\sixeufb=eufb6
\font\seveneufb=eufb7  \font\fiveeufb=eufb5
\font\teneurm=eurm10              
\font\eighteurm=eurm8 \font\nineeurm=eurm9
\font\teneurb=eurb10              
\font\eighteurb=eurb8 \font\nineeurb=eurb9
\font\teneusm=eusm10              
\font\eighteusm=eusm8 \font\nineeusm=eusm9
\font\teneusb=eusb10              
\font\eighteusb=eusb8 \font\nineeusb=eusb9
\else \def\eufm{\tt} \def\eufb{\tt} \def\eurm{\tt} \def\eurb{\tt}
\def\eusm{\tt} \def\eusb{\tt}    \fi}

\def\loadamsmath{\global\contaams=1 \ifx\arisposta\amsrisposta
\font\tenmsam=msam10 \font\ninemsam=msam9 \font\eightmsam=msam8
\font\sevenmsam=msam7 \font\sixmsam=msam6 \font\fivemsam=msam5
\font\tenmsbm=msbm10 \font\ninemsbm=msbm9 \font\eightmsbm=msbm8
\font\sevenmsbm=msbm7 \font\sixmsbm=msbm6 \font\fivemsbm=msbm5
\else \def\msbm{\bf} \fi \def\Bbb{\msbm} \def\symbl{\msam} \tenpoint}
\def\loadcyrill{\global\contacyrill=1 \ifx\arisposta\amsrisposta
\font\tenwncyr=wncyr10 \font\ninewncyr=wncyr9 \font\eightwncyr=wncyr8
\font\tenwncyb=wncyr10 \font\ninewncyb=wncyr9 \font\eightwncyb=wncyr8
\font\tenwncyi=wncyr10 \font\ninewncyi=wncyr9 \font\eightwncyi=wncyr8
\else \def\cyrill{\sl} \def\cyrilb{\sl} \def\cyrili{\sl} \fi\tenpoint}
\ifx\arisposta\amsrisposta
\font\sevenex=cmex7               
\font\eightex=cmex8  \font\nineex=cmex9
\font\ninecmmib=cmmib9   \font\eightcmmib=cmmib8
\font\sevencmmib=cmmib7 \font\sixcmmib=cmmib6
\font\fivecmmib=cmmib5   \skewchar\ninecmmib='177
\skewchar\eightcmmib='177  \skewchar\sevencmmib='177
\skewchar\sixcmmib='177   \skewchar\fivecmmib='177
\font\ninecmbsy=cmbsy9    \font\eightcmbsy=cmbsy8
\font\sevencmbsy=cmbsy7  \font\sixcmbsy=cmbsy6
\font\fivecmbsy=cmbsy5   \skewchar\ninecmbsy='60
\skewchar\eightcmbsy='60  \skewchar\sevencmbsy='60
\skewchar\sixcmbsy='60    \skewchar\fivecmbsy='60
\font\ninecmcsc=cmcsc9    \font\eightcmcsc=cmcsc8     \else
\def\cmmib{\fam\cmmibfam\tencmmib}\textfont\cmmibfam=\tencmmib
\scriptfont\cmmibfam=\tencmmib \scriptscriptfont\cmmibfam=\tencmmib
\def\cmbsy{\fam\cmbsyfam\tencmbsy} \textfont\cmbsyfam=\tencmbsy
\scriptfont\cmbsyfam=\tencmbsy \scriptscriptfont\cmbsyfam=\tencmbsy
\scriptfont\cmcscfam=\tencmcsc \scriptscriptfont\cmcscfam=\tencmcsc
\def\cmcsc{\fam\cmcscfam\tencmcsc} \textfont\cmcscfam=\tencmcsc \fi
%
\catcode`@=11
\newskip\ttglue
\gdef\tenpoint{\def\rm{\fam0\tenrm}
  \textfont0=\tenrm \scriptfont0=\sevenrm \scriptscriptfont0=\fiverm
  \textfont1=\teni \scriptfont1=\seveni \scriptscriptfont1=\fivei
  \textfont2=\tensy \scriptfont2=\sevensy \scriptscriptfont2=\fivesy
  \textfont3=\tenex \scriptfont3=\tenex \scriptscriptfont3=\tenex
  \def\mcal{\fam2 \tensy}  \def\mmit{\fam1 \teni}
  \textfont\itfam=\tenit \def\it{\fam\itfam\tenit}
  \textfont\slfam=\tensl \def\sl{\fam\slfam\tensl}
  \textfont\ttfam=\tentt \scriptfont\ttfam=\eighttt
  \scriptscriptfont\ttfam=\eighttt  \def\tt{\fam\ttfam\tentt}
  \textfont\bffam=\tenbf \scriptfont\bffam=\sevenbf
  \scriptscriptfont\bffam=\fivebf \def\bf{\fam\bffam\tenbf}
     \ifx\arisposta\amsrisposta    \ifnum\contaeuler=1
  \textfont\eufmfam=\teneufm \scriptfont\eufmfam=\seveneufm
  \scriptscriptfont\eufmfam=\fiveeufm \def\eufm{\fam\eufmfam\teneufm}
  \textfont\eufbfam=\teneufb \scriptfont\eufbfam=\seveneufb
  \scriptscriptfont\eufbfam=\fiveeufb \def\eufb{\fam\eufbfam\teneufb}
  \def\eurm{\teneurm} \def\eurb{\teneurb} \def\eusm{\teneusm}
  \def\eusb{\teneusb}    \fi    \ifnum\contaams=1
  \textfont\msamfam=\tenmsam \scriptfont\msamfam=\sevenmsam
  \scriptscriptfont\msamfam=\fivemsam \def\msam{\fam\msamfam\tenmsam}
  \textfont\msbmfam=\tenmsbm \scriptfont\msbmfam=\sevenmsbm
  \scriptscriptfont\msbmfam=\fivemsbm \def\msbm{\fam\msbmfam\tenmsbm}
     \fi      \ifnum\contacyrill=1     \def\cyrill{\tenwncyr}
  \def\cyrilb{\tenwncyb}  \def\cyrili{\tenwncyi}         \fi
  \textfont3=\tenex \scriptfont3=\sevenex \scriptscriptfont3=\sevenex
  \def\cmmib{\fam\cmmibfam\tencmmib} \scriptfont\cmmibfam=\sevencmmib
  \textfont\cmmibfam=\tencmmib  \scriptscriptfont\cmmibfam=\fivecmmib
  \def\cmbsy{\fam\cmbsyfam\tencmbsy} \scriptfont\cmbsyfam=\sevencmbsy
  \textfont\cmbsyfam=\tencmbsy  \scriptscriptfont\cmbsyfam=\fivecmbsy
  \def\cmcsc{\fam\cmcscfam\tencmcsc} \scriptfont\cmcscfam=\eightcmcsc
  \textfont\cmcscfam=\tencmcsc \scriptscriptfont\cmcscfam=\eightcmcsc
     \fi            \tt \ttglue=.5em plus.25em minus.15em
  \normalbaselineskip=12pt
  \setbox\strutbox=\hbox{\vrule height8.5pt depth3.5pt width0pt}
  \let\sc=\eightrm \let\big=\tenbig   \normalbaselines
  \baselineskip=\infralinea  \rm}
\gdef\ninepoint{\def\rm{\fam0\ninerm}
  \textfont0=\ninerm \scriptfont0=\sixrm \scriptscriptfont0=\fiverm
  \textfont1=\ninei \scriptfont1=\sixi \scriptscriptfont1=\fivei
  \textfont2=\ninesy \scriptfont2=\sixsy \scriptscriptfont2=\fivesy
  \textfont3=\tenex \scriptfont3=\tenex \scriptscriptfont3=\tenex
  \def\mcal{\fam2 \ninesy}  \def\mmit{\fam1 \ninei}
  \textfont\itfam=\nineit \def\it{\fam\itfam\nineit}
  \textfont\slfam=\ninesl \def\sl{\fam\slfam\ninesl}
  \textfont\ttfam=\ninett \scriptfont\ttfam=\eighttt
  \scriptscriptfont\ttfam=\eighttt \def\tt{\fam\ttfam\ninett}
  \textfont\bffam=\ninebf \scriptfont\bffam=\sixbf
  \scriptscriptfont\bffam=\fivebf \def\bf{\fam\bffam\ninebf}
     \ifx\arisposta\amsrisposta  \ifnum\contaeuler=1
  \textfont\eufmfam=\nineeufm \scriptfont\eufmfam=\sixeufm
  \scriptscriptfont\eufmfam=\fiveeufm \def\eufm{\fam\eufmfam\nineeufm}
  \textfont\eufbfam=\nineeufb \scriptfont\eufbfam=\sixeufb
  \scriptscriptfont\eufbfam=\fiveeufb \def\eufb{\fam\eufbfam\nineeufb}
  \def\eurm{\nineeurm} \def\eurb{\nineeurb} \def\eusm{\nineeusm}
  \def\eusb{\nineeusb}     \fi   \ifnum\contaams=1
  \textfont\msamfam=\ninemsam \scriptfont\msamfam=\sixmsam
  \scriptscriptfont\msamfam=\fivemsam \def\msam{\fam\msamfam\ninemsam}
  \textfont\msbmfam=\ninemsbm \scriptfont\msbmfam=\sixmsbm
  \scriptscriptfont\msbmfam=\fivemsbm \def\msbm{\fam\msbmfam\ninemsbm}
     \fi       \ifnum\contacyrill=1     \def\cyrill{\ninewncyr}
  \def\cyrilb{\ninewncyb}  \def\cyrili{\ninewncyi}         \fi
  \textfont3=\nineex \scriptfont3=\sevenex \scriptscriptfont3=\sevenex
  \def\cmmib{\fam\cmmibfam\ninecmmib}  \textfont\cmmibfam=\ninecmmib
  \scriptfont\cmmibfam=\sixcmmib \scriptscriptfont\cmmibfam=\fivecmmib
  \def\cmbsy{\fam\cmbsyfam\ninecmbsy}  \textfont\cmbsyfam=\ninecmbsy
  \scriptfont\cmbsyfam=\sixcmbsy \scriptscriptfont\cmbsyfam=\fivecmbsy
  \def\cmcsc{\fam\cmcscfam\ninecmcsc} \scriptfont\cmcscfam=\eightcmcsc
  \textfont\cmcscfam=\ninecmcsc \scriptscriptfont\cmcscfam=\eightcmcsc
     \fi            \tt \ttglue=.5em plus.25em minus.15em
  \normalbaselineskip=11pt
  \setbox\strutbox=\hbox{\vrule height8pt depth3pt width0pt}
  \let\sc=\sevenrm \let\big=\ninebig \normalbaselines\rm}
\gdef\eightpoint{\def\rm{\fam0\eightrm}
  \textfont0=\eightrm \scriptfont0=\sixrm \scriptscriptfont0=\fiverm
  \textfont1=\eighti \scriptfont1=\sixi \scriptscriptfont1=\fivei
  \textfont2=\eightsy \scriptfont2=\sixsy \scriptscriptfont2=\fivesy
  \textfont3=\tenex \scriptfont3=\tenex \scriptscriptfont3=\tenex
  \def\mcal{\fam2 \eightsy}  \def\mmit{\fam1 \eighti}
  \textfont\itfam=\eightit \def\it{\fam\itfam\eightit}
  \textfont\slfam=\eightsl \def\sl{\fam\slfam\eightsl}
  \textfont\ttfam=\eighttt \scriptfont\ttfam=\eighttt
  \scriptscriptfont\ttfam=\eighttt \def\tt{\fam\ttfam\eighttt}
  \textfont\bffam=\eightbf \scriptfont\bffam=\sixbf
  \scriptscriptfont\bffam=\fivebf \def\bf{\fam\bffam\eightbf}
     \ifx\arisposta\amsrisposta   \ifnum\contaeuler=1
  \textfont\eufmfam=\eighteufm \scriptfont\eufmfam=\sixeufm
  \scriptscriptfont\eufmfam=\fiveeufm \def\eufm{\fam\eufmfam\eighteufm}
  \textfont\eufbfam=\eighteufb \scriptfont\eufbfam=\sixeufb
  \scriptscriptfont\eufbfam=\fiveeufb \def\eufb{\fam\eufbfam\eighteufb}
  \def\eurm{\eighteurm} \def\eurb{\eighteurb} \def\eusm{\eighteusm}
  \def\eusb{\eighteusb}       \fi    \ifnum\contaams=1
  \textfont\msamfam=\eightmsam \scriptfont\msamfam=\sixmsam
  \scriptscriptfont\msamfam=\fivemsam \def\msam{\fam\msamfam\eightmsam}
  \textfont\msbmfam=\eightmsbm \scriptfont\msbmfam=\sixmsbm
  \scriptscriptfont\msbmfam=\fivemsbm \def\msbm{\fam\msbmfam\eightmsbm}
     \fi       \ifnum\contacyrill=1     \def\cyrill{\eightwncyr}
  \def\cyrilb{\eightwncyb}  \def\cyrili{\eightwncyi}         \fi
  \textfont3=\eightex \scriptfont3=\sevenex \scriptscriptfont3=\sevenex
  \def\cmmib{\fam\cmmibfam\eightcmmib}  \textfont\cmmibfam=\eightcmmib
  \scriptfont\cmmibfam=\sixcmmib \scriptscriptfont\cmmibfam=\fivecmmib
  \def\cmbsy{\fam\cmbsyfam\eightcmbsy}  \textfont\cmbsyfam=\eightcmbsy
  \scriptfont\cmbsyfam=\sixcmbsy \scriptscriptfont\cmbsyfam=\fivecmbsy
  \def\cmcsc{\fam\cmcscfam\eightcmcsc} \scriptfont\cmcscfam=\eightcmcsc
  \textfont\cmcscfam=\eightcmcsc \scriptscriptfont\cmcscfam=\eightcmcsc
     \fi             \tt \ttglue=.5em plus.25em minus.15em
  \normalbaselineskip=9pt
  \setbox\strutbox=\hbox{\vrule height7pt depth2pt width0pt}
  \let\sc=\sixrm \let\big=\eightbig \normalbaselines\rm }
\gdef\tenbig#1{{\hbox{$\left#1\vbox to8.5pt{}\right.\n@space$}}}
\gdef\ninebig#1{{\hbox{$\textfont0=\tenrm\textfont2=\tensy
   \left#1\vbox to7.25pt{}\right.\n@space$}}}
\gdef\eightbig#1{{\hbox{$\textfont0=\ninerm\textfont2=\ninesy
   \left#1\vbox to6.5pt{}\right.\n@space$}}}
\def\alternativefont#1#2{\ifx\arisposta\amsrisposta \relax \else
\xdef#1{#2} \fi}
\global\contaeuler=0 \global\contacyrill=0 \global\contaams=0
%
%
%
\newbox\fotlinebb \newbox\hedlinebb \newbox\leftcolumn
\gdef\makeheadline{\vbox to 0pt{\vskip-22.5pt
     \fullline{\vbox to8.5pt{}\the\headline}\vss}\nointerlineskip}
\gdef\makehedlinebb{\vbox to 0pt{\vskip-22.5pt
     \fullline{\vbox to8.5pt{}\copy\hedlinebb\hfil
     \line{\hfill\the\headline\hfill}}\vss} \nointerlineskip}
\gdef\makefootline{\baselineskip=24pt \fullline{\the\footline}}
\gdef\makefotlinebb{\baselineskip=24pt
    \fullline{\copy\fotlinebb\hfil\line{\hfill\the\footline\hfill}}}
\gdef\doubleformat{\shipout\vbox{\Landspec\makehedlinebb
     \fullline{\box\leftcolumn\hfil\columnbox}\makefotlinebb}
     \advancepageno}
\gdef\columnbox{\leftline{\pagebody}}
\gdef\line#1{\hbox to\hsize{\hskip\leftskip#1\hskip\rightskip}}
\gdef\fullline#1{\hbox to\fullhsize{\hskip\leftskip{#1}%
\hskip\rightskip}}
\gdef\footnote#1{\let\@sf=\empty
         \ifhmode\edef\#sf{\spacefactor=\the\spacefactor}\/\fi
         #1\@sf\vfootnote{#1}}
\gdef\vfootnote#1{\insert\footins\bgroup
         \ifnum\dimnota=1  \eightpoint\fi
         \ifnum\dimnota=2  \ninepoint\fi
         \ifnum\dimnota=0  \tenpoint\fi
         \interlinepenalty=\interfootnotelinepenalty
         \splittopskip=\ht\strutbox
         \splitmaxdepth=\dp\strutbox \floatingpenalty=20000
         \leftskip=\oldssposta \rightskip=\olddsposta
         \spaceskip=0pt \xspaceskip=0pt
         \ifnum\sinnota=0   \textindent{#1}\fi
         \ifnum\sinnota=1   \item{#1}\fi
         \footstrut\futurelet\next\fo@t}
\gdef\fo@t{\ifcat\bgroup\noexpand\next \let\next\f@@t
             \else\let\next\f@t\fi \next}
\gdef\f@@t{\bgroup\aftergroup\@foot\let\next}
\gdef\f@t#1{#1\@foot} \gdef\@foot{\strut\egroup}
\gdef\footstrut{\vbox to\splittopskip{}}
\skip\footins=\bigskipamount
\count\footins=1000  \dimen\footins=8in
\catcode`@=12
\tenpoint
\ifnum\unoduecol=1 \hsize=\tothsize   \fullhsize=\tothsize \fi
\ifnum\unoduecol=2 \hsize=\collhsize  \fullhsize=\tothsize \fi
\global\let\lrcol=L
\ifnum\unoduecol=1 \output{\plainoutput{\ifnum\tipbnota=2
\clearnmbnota\fi}} \fi
\ifnum\unoduecol=2 \output{\if L\lrcol
     \global\setbox\leftcolumn=\columnbox
     \global\setbox\fotlinebb=\line{\hfill\the\footline\hfill}
     \global\setbox\hedlinebb=\line{\hfill\the\headline\hfill}
     \advancepageno  \global\let\lrcol=R
     \else  \doubleformat \global\let\lrcol=L \fi
     \ifnum\outputpenalty>-20000 \else\dosupereject\fi
     \ifnum\tipbnota=2\clearnmbnota\fi }\fi
\def\ifdoublepage{\ifnum\unoduecol=2 }
\gdef\yespagenumbers{\footline={\hss\tenrm\folio\hss}}
\gdef\ciao{\par\vfill\supereject \ifnum\unoduecol=2
     \if R\lrcol  \headline={}\nopagenumbers\null\vfill\eject
     \fi\fi \end}

\newskip\olddsposta \newskip\oldssposta
\global\oldssposta=\leftskip \global\olddsposta=\rightskip

\def\filldots{\leaders\hbox to 1em{\hss.\hss}\hfill}
\def\inquadrb#1 {\vbox {\hrule  \hbox{\vrule \vbox {\vskip .2cm
    \hbox {\ #1\ } \vskip .2cm } \vrule  }  \hrule} }
 \def\newline{\hfil\break}
\def\jump{\vskip\baselineskip} \newskip\iinnffrr
\def\sjump{\iinnffrr=\baselineskip
          \divide\iinnffrr by 2 \vskip\iinnffrr}
\def\bjump{\vskip\baselineskip \vskip\baselineskip}
\newcount\nmbnota  \def\clearnmbnota{\global\nmbnota=0}
\newcount\tipbnota \def\letterfootnote{\global\tipbnota=1}

\def\note#1{\global\advance\nmbnota by 1 \ifnum\tipbnota=1
    \footnote{$^{\rm\nttlett}$}{#1} \else {\ifnum\tipbnota=2
    \footnote{$^{\nttsymb}$}{#1}
    \else\footnote{$^{\the\nmbnota}$}{#1}\fi}\fi}
\def\nttlett{\ifcase\nmbnota \or a\or b\or c\or d\or e\or f\or
g\or h\or i\or j\or k\or l\or m\or n\or o\or p\or q\or r\or
s\or t\or u\or v\or w\or y\or x\or z\fi}
\def\nttsymb{\ifcase\nmbnota \or\dag\or\sharp\or\ddag\or\star\or
\natural\or\flat\or\clubsuit\or\diamondsuit\or\heartsuit
\or\spadesuit\fi}   \clearnmbnota
\def\numberfootnote{\global\tipbnota=0} \numberfootnote
\def\setnote#1{\expandafter\xdef\csname#1\endcsname{
\ifnum\tipbnota=1 {\rm\nttlett} \else {\ifnum\tipbnota=2
{\nttsymb} \else \the\nmbnota\fi}\fi} }
\newcount\nbmfig  \def\clearnbmfig{\global\nbmfig=0}
\gdef\figure{\global\advance\nbmfig by 1
      {\rm fig. \the\nbmfig}}   \clearnbmfig
\def\setfig#1{\expandafter\xdef\csname#1\endcsname{fig. \the\nbmfig}}

\newcount\frmcount \def\clearfrmcount{\global\frmcount=0}
\def\numero{\global\advance\frmcount by 1   \ifnum\indappcount=0
  {\ifnum\cpcount <1 {\hbox{\rm (\the\frmcount )}}  \else
  {\hbox{\rm (\the\cpcount .\the\frmcount )}} \fi}  \else
  {\hbox{\rm (\applett .\the\frmcount )}} \fi}
\def\nameformula#1{\global\advance\frmcount by 1%
\ifnum\draftnum=0  {\ifnum\indappcount=0%
{\ifnum\cpcount<1\xdef\spzzttrra{(\the\frmcount )}%
\else\xdef\spzzttrra{(\the\cpcount .\the\frmcount )}\fi}%
\else\xdef\spzzttrra{(\applett .\the\frmcount )}\fi}%
\else\xdef\spzzttrra{(#1)}\fi%
\expandafter\xdef\csname#1\endcsname{\spzzttrra}
\eqno \ifnum\draftnum=0 {\ifnum\indappcount=0
  {\ifnum\cpcount <1 {\hbox{\rm (\the\frmcount )}}  \else
  {\hbox{\rm (\the\cpcount .\the\frmcount )}}\fi}   \else
  {\hbox{\rm (\applett .\the\frmcount )}} \fi} \else (#1) \fi $$}
\def\nfr{\nameformula}    
\def\nameali#1{\global\advance\frmcount by 1%
\ifnum\draftnum=0  {\ifnum\indappcount=0%
{\ifnum\cpcount<1\xdef\spzzttrra{(\the\frmcount )}%
\else\xdef\spzzttrra{(\the\cpcount .\the\frmcount )}\fi}%
\else\xdef\spzzttrra{(\applett .\the\frmcount )}\fi}%
\else\xdef\spzzttrra{(#1)}\fi%
\expandafter\xdef\csname#1\endcsname{\spzzttrra}
  \ifnum\draftnum=0  {\ifnum\indappcount=0
  {\ifnum\cpcount <1 {\hbox{\rm (\the\frmcount )}}  \else
  {\hbox{\rm (\the\cpcount .\the\frmcount )}}\fi}   \else
  {\hbox{\rm (\applett .\the\frmcount )}} \fi} \else (#1) \fi}
\clearfrmcount
\newcount\cpcount \def\clearcpcount{\global\cpcount=0}
\newcount\subcpcount \def\clearsubcpcount{\global\subcpcount=0}
\newcount\appcount \def\clearappcount{\global\appcount=0}
\newcount\indappcount \def\clearindappcount{\indappcount=0}
\newcount\sottoparcount 

\def\applett{\ifcase\appcount  \or {A}\or {B}\or {C}\or
{D}\or {E}\or {F}\or {G}\or {H}\or {I}\or {J}\or {K}\or {L}\or
{M}\or {N}\or {O}\or {P}\or {Q}\or {R}\or {S}\or {T}\or {U}\or
{V}\or {W}\or {X}\or {Y}\or {Z}\fi
             \ifnum\appcount<0
    \message{>>  ERROR: counter \appcount out of range <<}\fi
             \ifnum\appcount>26
   \message{>>  ERROR: counter \appcount out of range <<}\fi}
\clearappcount  \clearindappcount
\newcount\connttrre  \def\clearconnttrre{\global\connttrre=0}
\newcount\countref  \def\clearcountref{\global\countref=0}
\clearcountref
\def\chapter#1{\global\advance\cpcount by 1 \clearfrmcount
                 \goodbreak\null\vbox{\jump\nobreak
                 \clearsubcpcount\clearindappcount
                 \itemitem{\ttaarr\the\cpcount .\qquad}{\ttaarr #1}
                 \par\nobreak\jump\sjump}\nobreak}
\def\section#1{\global\advance\subcpcount by 1 \goodbreak\null
               \vbox{\sjump\nobreak\ifnum\indappcount=0
                 {\ifnum\cpcount=0 {\itemitem{\ppaarr
               .\the\subcpcount\quad\enskip\ }{\ppaarr #1}\par} \else
                 {\itemitem{\ppaarr\the\cpcount .\the\subcpcount\quad
                  \enskip\ }{\ppaarr #1} \par}  \fi}
                \else{\itemitem{\ppaarr\applett .\the\subcpcount\quad
                 \enskip\ }{\ppaarr #1}\par}\fi\nobreak\jump}\nobreak}
\clearsubcpcount
\def\appendix#1{\global\advance\appcount by 1 \clearfrmcount
                  \goodbreak\null\vbox{\jump\nobreak
                  \global\advance\indappcount by 1 \clearsubcpcount
                  \itemitem{\ttaarr App.\applett\ }{\ttaarr #1}
                  \nobreak\jump\sjump}\nobreak}
\clearappcount \clearindappcount

\clearcpcount\clearcountref

\def\setchap#1{\ifnum\indappcount=0{\ifnum\subcpcount=0%
\xdef\spzzttrra{\the\cpcount}%
\else\xdef\spzzttrra{\the\cpcount .\the\subcpcount}\fi}
\else{\ifnum\subcpcount=0 \xdef\spzzttrra{\applett}%
\else\xdef\spzzttrra{\applett .\the\subcpcount}\fi}\fi
\expandafter\xdef\csname#1\endcsname{\spzzttrra}}
\newcount\draftnum \newcount\ppora   \newcount\ppminuti
\global\ppora=\time   \global\ppminuti=\time
\global\divide\ppora by 60  \draftnum=\ppora
\multiply\draftnum by 60    \global\advance\ppminuti by -\draftnum
\global\draftnum=0
\def\droggi{\number\day /\number\month /\number\year\ \the\ppora
:\the\ppminuti}
 \global\draftnum=0
\def\draftcomment#1{\ifnum\draftnum=0 \relax \else
{\ {\bf ***}\ #1\ {\bf ***}\ }\fi} 
%
%
\catcode`@=11
\gdef\Ref#1{\expandafter\ifx\csname @rrxx@#1\endcsname\relax%
{\global\advance\countref by 1%
\ifnum\countref>200%
\message{>>> ERROR: maximum number of references exceeded <<<}%
\expandafter\xdef\csname @rrxx@#1\endcsname{0}\else%
\expandafter\xdef\csname @rrxx@#1\endcsname{\the\countref}\fi}\fi%
\ifnum\draftnum=0 \csname @rrxx@#1\endcsname \else#1\fi}
\gdef\beginref{\ifnum\draftnum=0  \gdef\Rref{\fairef}
\gdef\endref{\scriviref} \else\relax\fi
\ifx\risposta\mplarisposta \ninepoint \fi
\parskip 2pt plus 2pt \baselineskip=12pt}
\def\Reflab#1{[#1]} \gdef\Rref#1#2{\item{\Reflab{#1}}{#2}}
\gdef\endref{\relax}  \newcount\conttemp
\gdef\fairef#1#2{\expandafter\ifx\csname @rrxx@#1\endcsname\relax
{\global\conttemp=0
\message{>>> ERROR: reference [#1] not defined <<<} } \else
{\global\conttemp=\csname @rrxx@#1\endcsname } \fi
\global\advance\conttemp by 50
\global\setbox\conttemp=\hbox{#2} }
\gdef\scriviref{\clearconnttrre\conttemp=50
\loop\ifnum\connttrre<\countref \advance\conttemp by 1
\advance\connttrre by 1
\item{\Reflab{\the\connttrre}}{\unhcopy\conttemp} \repeat}
\clearcountref \clearconnttrre
\catcode`@=12
\ifx\risposta\mplarisposta \def\Reflab#1{#1.} \letterfootnote \fi

\def\slashchar#1{\setbox0=\hbox{$#1$} \dimen0=\wd0
     \setbox1=\hbox{/} \dimen1=\wd1 \ifdim\dimen0>\dimen1
      \rlap{\hbox to \dimen0{\hfil/\hfil}} #1 \else
      \rlap{\hbox to \dimen1{\hfil$#1$\hfil}} / \fi}
\ifx\oldchi\undefined \let\oldchi=\chi
  \def\cchi{{\raise 1pt\hbox{$\oldchi$}}} \let\chi=\cchi \fi

\def\frac#1#2{{\textstyle{#1 \over #2}}}

\def\half{\ifinner {\scriptstyle {1 \over 2}}\else {1 \over 2} \fi}

\def\simge{\rlap{\raise 2pt \hbox{$>$}}{\lower 2pt \hbox{$\sim$}}}
\def\simle{\rlap{\raise 2pt \hbox{$<$}}{\lower 2pt \hbox{$\sim$}}}

\def\vbig#1#2{{\vbigd@men=#2\divide\vbigd@men by 2%
\hbox{$\left#1\vbox to \vbigd@men{}\right.\n@space$}}}

\null
%
%
%
%
\nopagenumbers{\baselineskip=12pt
\line{\hfill CBPF-NF-064/94}
\ifdoublepage \bjump\bjump\bjump\bjump\else\vfill\fi
\centerline{\capsone Sp(2) Covariant Quantisation of General}
\sjump
\centerline{\capsone Gauge Theories.}
\bjump
\centerline{\scaps {Jos\'e-Luis V\'azquez-Bello}}
\sjump
\centerline{\sl CBPF-CNPq/CLAF Centro Brasileiro de Pesquisas Fisicas}
\centerline{\sl Rua Dr. Xavier Sigaud 150, CEP. 22290}
\centerline{\sl  Rio de  Janeiro - RJ, BRASIL.}
\sjump
\centerline{ {\sl e-mail:} bello@cbpfsu7.cat.cbpf.br}
\vfill
\ifnum\unoduecol=2 \eject\null\vfill\fi
\centerline{\capsone abstract }
\sjump
\noindent
{This letter studies the Sp(2) covariant quantisation of gauge theories.
The geometrical interpretation of gauge theories in terms of quasi principal
fibre bundles $Q(M_S ,G_S)$ is reviewed. It is then described the Sp(2)
algebra of ordinary Yang-Mills theory. A consistent formulation of covariant
lagrangian quantisation for general gauge theories based on Sp(2) BRST
symmetry is established. The original $N=1$, ten dimensional superparticle
is considered as an example of infinitely reducible gauge algebras,
and given explicitly its Sp(2) BRST invariant action. }

\sjump
\ifnum\unoduecol=2 \vfill\fi
\eject
\yespagenumbers\pageno=1

\def\scm{M}
\def\scn{N}
\def\scmn{MN}

\def\cala{\cal A}
\def\half{{1\over2}}
\def\twelfe{{1\over {12}}}

\def\delmu{\partial_\mu}
\def\xmu {x^\mu}
\def\pmu {p_\mu}
\def\Amu {A_\mu}
\def\gammu {\gamma^\mu}
\def\psh {\rlap{/}{p}}
\def\thetaa {{\theta^a}}
\def\thetab {{\theta^b}}
\def\thetac {{\theta^c}}

\def\ginv {g^{-1}}
\def\varri { {\delta_r}\over {\delta\Phi^A} }
\def\varle { {\delta_l}\over {\delta\Phi^\star_A} }
\def\varrid {{ {\delta  }\over {\delta\Phi^A} }}
\def\varlea {{ {\delta_l}\over {\delta\Phi^\star_{Aa}} }}
\def\varrist  {{ {\delta_r}\over {\delta\Phi^{\star\star}_A} }}
\def\notequiv{\rlap{/}{=}}

\chapter {Gauge theories in terms of quasi-principal fibre bundles.}

Gauge theories have a nice geometrical interpretation in terms of connections
on a principal fibre bundle (pfb) $P(M,G)$, where $M$ is the base space-time
manifold and G is the gauge group [\Ref{bon1},\Ref{bon2}, \Ref{hoyos},
\Ref{rogers}].
However, quantisation of gauge theories requires the introduction
of fields $(c^n_m ,\pi^n_m)$. It would be then desirable to have
a formalism where those extra fields fit into some representation of a
larger group and all the fields are components of a superfield.
This is a step in the direction of recovering a geometrical interpretation
of quantum gauge theories. The main ingredients in the construction of
geometrical quasi-principal fibre bundles
(qpfb) are a space-time base manifold $M$, a gauge group $G$,
an extended superspace manifold $M_S$ which is obtained by adding
two extra Grassmann variables $\theta^a$ $(a=1,2)$ to $M$, in the case of
$Sp(2)$ symmetry, and a supergroup $G_S$.
The construction is performed basically in three steps
[\Ref{bon1},\Ref{bon2}, \Ref{hoyos}].
It starts with a pfb $P(M,G)$ and extend the gauge group $G$
to a supergroup $G_S$. The composition of $G$ with a Grassmann algebra $B$
prolongs $P(M,G)$ to a pfb $P'(M,G_S )$.
The most general supergroup $G_S$ can be represented in matrix form. In
particular, $OSp(N/M)$ groups are represented by block matrices of the form
$$
\pmatrix { A & E \cr
           C & D \cr}
\nfr{matrix}
where
$A$, $D$ are $(N\times N)$ and $(M\times M)$ matrices whose elements
are taken from the even part of the Grassmann algebra $B$ constructed
over a complex vector space $W$, whilst $E$, $C$ are $(N\times M)$ and
$(N\times M)$ rectangular matrices whose elements belong to the
odd part of $B$.
Next, it is enlarged the base space manifold $M$ to a superspace $M_S$ in
$P'(M,G_S )$ by adding Grassmann variables. At this stage, a pfb
$P''(M_S ,G_S )$ is obtained. Finally, the pfb $P'' (M_S G_S )$ is transform
into a quasi-principal fibre bundle $Q(M_S ,G_S )$.
For instance,
given a one-form valued function $\alpha (x) = \Amu d\xmu$ on $M$ this
induces a connection $\omega$ on the pfb $P(M,G)$. Then a one-form valued
function $\alpha '$ on $M_S$ is found by
$$
\alpha ' (x,\thetaa )= \ginv\Amu d\xmu g + \ginv dg
\nfr{qalpha}
where
$g = g(\xmu ,\thetaa )$, $(a=1,2,\dots )$ which
induces a connection $\omega '$ on the qpfb $Q(M_S,G_S)$
[\Ref{bon1},\Ref{bon2},\Ref{hoyos},\Ref{rogers}].

\chapter{The $Sp(2)$ BRST Algebra of Yang-Mills Theory.}

It has
been realized for some time [\Ref{baul},\Ref{our},\Ref{tieg},\Ref{mieg}] that
a geometrical construction can be useful for the discussion of BRST and
anti-BRST symmetry.
The idea is to use a superspace with coordinates
$Z^M = (x^\mu ,\theta^a)$, where $(a=1,2)$ and $\theta^a$ is an
anti-commuting scalar coordinate
and the BRST generators $s^a$ are realized as differential
operators on superspace,
$s^a={\partial\over{\partial\thetaa}}$, so that $s^a s^b + s^b s^a =0$
holds automatically
\footnote{$^1$}
{It
is usually defined a bosonic operator
$\sigma = \half\epsilon_{ab} s^a s^b$ where $\epsilon_{ab}$ is
the symplectic invariant form of $Sp(2)$, so that
$\epsilon_{ab} = -\epsilon_{ba}$,  $\epsilon^{ab}\epsilon_{bc}=\delta_{ac}$
and $\epsilon_{12}=1$. The generator $\sigma$ is invariant under  $Sp(2)$
and satisfies $s^a \sigma = 0 $. The Sp(2) generators
$\sigma^i$ $(i=\pm ,0)$ and the fermionic charges
$s^a$ together form an algebra which is a contraction of
$OSp(1, 1/2)$ and denoted as $ISp(2)$
[\Ref{espies}].}. For example, in
Yang-Mills theory the gauge potential $A^i_\mu$ and the Faddeev-Popov
ghost $(c^a)^i$ (where $i$ is an adjoint group index) can be
combined into a super-gauge field ${\cal A}_M^i (Z)$ whose lowest
order components are
${\cal A}_{\scm}^i (Z)\big\vert_{\thetaa =0}
=({\cal A}_\mu^i ,{\cal A}^i_\thetaa )\big\vert_{\thetaa =0}
=(A_\mu^i ,{c^a}^i )$.
Then the standard Yang-Mills BRST transformations arise
from imposing the constraints
${\cal F}_{\mu\thetaa}^i=0$, ${\cal F}^i_{\thetaa\thetab}=0$
on the superfield strength ${\cal F}^i_{\scmn}$ [\Ref{baul},\Ref{our}].
This gives an elegant geometrical description of BRST and anti-BRST symmetry.

Let us review the construction
of gauge theories in the superspace with
coordinates $Z^{\scm}=(x^\mu ,\thetaa )$, which gives a geometric
formulation of $Sp(2)$ BRST symmetry.
We consider matter fields $\Phi^i (x,\thetaa )$ and
a gauge potential
${\cal A}^i_{\scm}(x,\thetaa )
=\bigl ({\cal A}^i_\mu (x,\thetaa ),{\cal A}^i_\thetaa (x,\thetaa )\bigl )$.
These can be used to  define a covariant derivative
$$
{\cal D}_{\scm}\Phi^i
=\partial_{\scm}\Phi^i -{(T^k)^i}_j{\cal A}^k_{\scm}\Phi^j
\nfr{covariant}
and the field strength
$${\cal F}^i_{\scmn}
=\partial_{\scm}{\cal A}^i_{\scn}
-(-1)^{\scmn}\partial_{\scn}{\cal A}^i_{\scm}
+{f^i}_{jk}{\cal A}^j_{\scm}{\cal A}^k_{\scn}
\nfr{strength}
where
$(-1)^{\scmn}$ is 1 unless both $M$ and $N$ are indices referring
to anti-commuting coordinates, in which case it is $-1$.
The gauge potential ${\cal A}_{\scm}$ contains more
component fields than the physical gauge and ghost fields
and so, as in supersymmetric theories, constraints should be imposed
on the field strength ${\cal F}$. Appropriate constraints are
[\Ref{baul},\Ref{our}]
$${\cal F}_{\thetaa\thetab} =0,\qquad {\cal F}_{\mu\thetaa} =0.
\nfr{constraints}
These
can be written
more explicitly
as
$$\delmu{\cala}_\thetaa -\partial_\thetaa {\cala}_\mu
                     +[{\cala}_\mu ,{\cala}_\thetaa ]=0\nfr{acala}
$$\partial_\thetaa{\cala}_\thetaa
           +{1\over 2}[{\cala}_\thetaa ,{\cala}_\thetaa ]=0\nfr{bcala}
$$\epsilon^{ab}\Big(\partial_\thetaa {\cala}_\thetab
      + [{\cala}_\thetaa,{\cala}_\thetab ]\Big) =0.
\nfr{ecala}
Defining
the component expansions
$${\cala}_\mu (x,\thetaa )
             =A_\mu (x)+\thetaa \Lambda_{a\mu} (x)
                           +\thetaa\thetab \Omega_{ab\mu} (x)\nfr{aexp}
$${\cala}^b_\thetaa (x,\thetaa )
         =c^b (x) + \thetaa \Upsilon^b_a (x) + \thetaa\thetac \omega^b_{ac}
(x),\nfr{thexp}
however, it was found that
if $A_\mu$, $c^a$, $\pi$ are identified with the gauge, ghost
(anti-ghost) and auxiliary fields respectively then the supergauge fields
have the expansions
\footnote{$^\star$}
{In Ref [\Ref{our}], it was obtained explicitly a geometrical formulation
of BRST and anti-BRST symmetries and given the field content of
$\Lambda$, $\Upsilon$, $\Omega$ and $\omega$. The components in the
expansion can also be read as conditions on the mapping of the coordinates
$\phi^i$ of the fibres over $\{ {\cal U}_i \}$ (covering set of $M_G$) and
expressed as cocycle conditions.}
$$
{\cala}_\mu = A_\mu + \thetaa (s^a A_\mu ) + \thetaa\thetab (s^a s^b A_\mu )
\nfr{aexpan}
$${\cala}^b_\thetaa = c^b + \thetaa (s^a c^b ) + \thetaa\thetac (s^a s^c c^b )
.\nfr{thexpan}
The
BRST and anti-BRST generators $s^a$ are
then identified with the superspace differential operators $\partial_\thetaa$
[\Ref{baul},\Ref{our}], and the complete set of BRST and anti-BRST
transformations are given by [\Ref{ore} ,\Ref{gaume} ,\Ref{spirid} ]
$$
\eqalign{
s^a \phi^i &= R^i_\alpha c^{\alpha a}, \qquad
s^a c^{\alpha b} =\epsilon^{ab} \pi^\alpha - \half f^\alpha_{\beta\gamma}
c^{\beta a} c^{\gamma b}, \cr
s^a \pi^\alpha &= \half f^\alpha_{\beta\gamma}\pi^\beta c^{\gamma a}
-\twelfe (f^\alpha_{\beta\gamma} f^\beta_{\delta\tau} +
f^\alpha_{\delta\tau ,i} R^i_\gamma )c^{\delta a}c^{\tau e}
\epsilon_{eb}c^{\gamma b},\cr}
\nfr{setbrst}
where
the generators $R^i_\alpha$ for the gauge field $A^i_\mu$
read off from $s^a A^i_\mu = (D_\mu c^a )^i$, and $\pi^\alpha$
is an auxiliary field which connects ghost and antighost sectors.
%
%

For
the matter fields $\Phi^i (x,\thetaa )$, we impose the constraint
$${\cal D}_\thetaa\Phi^i
     =\partial_\thetaa \Phi^i - {(T^k )^i}_j{\cal A}^k_\thetaa\Phi^j = 0
\nfr{constrmatter}
which
implies
$$\Phi^i = \psi^i + \thetaa (s^a \psi^i ),\nfr{matterx}
and the BRST and anti-BRST transformations again corresponds to
translations in the $\thetaa$ direction with
$\partial\over{\partial\thetaa}$ realized as differential operators
on the extended space manifold $M_S$.

\chapter{Sp(2) formalism for General gauge theories.}

Consider a
general gauge theory with classical fields $A^i (\xmu)$
$(i=1,2,\dots ,n)$ and classical action $S_0 (A^i )$. The action is
invariant under gauge transformations
$$
\delta A^i = R^i_\alpha \xi^\alpha,\nfr{gaugeinv}
where
$\xi^\alpha$ is the local gauge parameter.
The Noether equations are given by
$$
S_0,i \;\; R^i_\alpha = 0 ,\qquad \alpha =1,2,\dots ,m\qquad (0\leq m\leq n),
\nfr{noether}
and
the generators of the gauge transformations satisfy
$$
R^i_{\alpha ,j} R^j_\beta - (-)^{\alpha\beta} R^i_{\beta ,j}R^j_{\alpha}
   =-R^i_{\gamma} {f^\gamma}_{\alpha\beta} - S_{0,j} M^{ij}_{\alpha\beta}.
\nfr{generat}

To construct a covariant lagrangian formalism for general gauge theories
either with open ($M^{ij}_{\alpha\beta} \notequiv 0$) or closed algebras, and
based on $Sp(2)$ BRST symmetry, it is needed to enlarge the base manifold
$M$ to $M_S$. It is then defined a superspace $M_S$ to include classical
fields
$\Phi^A$ and $Sp(2)$ doublets of anti-fields $\Phi^\star_A$,
$\Phi^{\star\star}_A$
[\Ref{lavrovx},\Ref{lavrov},\Ref{hull}].
The properties of these fields and anti-fields are
$\epsilon (\Phi^A ) =\epsilon_A$, $\epsilon (\Phi^\star_A ) =\epsilon_A + 1$,
$\epsilon (\Phi^{\star\star}_A ) =\epsilon_A$,
$gh (\Phi^\star_A ) = (-)^a - gh (\Phi^A )$ and
$gh (\Phi^{\star\star}_A ) = - gh (\Phi^A )$. An extended Poisson
superbracket is defined by
$$
(F,G) = {\varri}F{\varle}G -(-)^{\epsilon_F\epsilon_G} {\varri}G{\varle}F,
\nfr{poisson}
where
${\epsilon_F}$, ${\epsilon_G}$ denotes the Grassmann parity of the $F$, $G$
functions on $M_S$, and left (right) derivatives are understood
with respect to anti-fields (fields) unless otherwise stated.
The extended anti-bracket
\poisson\ satisfies
$$
\epsilon\Big( (F,G)\Big) = \epsilon (F) + \epsilon (G) + 1
\nfr{propone}
$$gh \Big( (F,G)^a\Big) = - (-)^a + gh (F) + gh (G), \qquad a=1,2
\nfr{proptwo}
$$ (F,G) = - (-)^{\epsilon_F\epsilon_G} (G,F) \nfr{propthree}
and
$$(-)^{\epsilon_F}{\epsilon_G} \big( (F,G), H\big) + [cycl.\;perm\;\;
(F,G,H)] = 0. \nfr{propfour}

A bosonic
action functional $S = S(\Phi^A ,\Phi^\star_A ,\Phi^{\star\star}_A )$
is constructed on $M_S$. This action satisfy the following {\it generating
equation}
$$
\bar\Delta^a\exp^{{i\over\hbar}
S(\Phi^A ,\Phi^\star_A ,\Phi^{\star\star}_A )} = 0, \qquad  (a=1,2),
\nfr{generating}
together
with the boundary condition
$$
S(\Phi^A ,\Phi^\star_A ,\Phi^{\star\star}_A )
\Big\vert_{\Phi^\star_A =\Phi^{\star\star}_A =0} = S (\Phi_A ).
\nfr{boundary}
The
operator $\bar\Delta^a$ is defined by
$$
\bar\Delta^a = \Delta^a  + (i/\hbar ) V^a, \qquad
\Delta^a = (-)^{\epsilon_A} \varrid \varlea
\qquad
V^a = \epsilon^{ab}\Phi^\star_{Ab} \varrist .
\nfr{operators}
The
algebra of operators \operators\ satisfy the important property
\footnote{$^2$}
{A supercommutative, associative algebra $\cal A$ equipped with an extended
Poisson anti-bracket structure plus a nilpotent property it is known
as a BV-algebra, or coboundary Gersterhaber algebra (CGA) [\Ref{galgebra}].}
$$
\bar\Delta^{\{a} \bar\Delta^{b\}} = 0.
\nfr{nilpotency}
The solution
to the generating equation \generating\ is given as a power series
of the Planck constant
$$
S(\Phi^A ,\Phi^\star_A ,\Phi^{\star\star}_A ) = \sum^{+\infty}_{n=0}
{\hbar}^n S_{(n)},\nfr{power}
where
the classical approximation $S_{(0)}$ satisfies
$$
\half ( S_{(0)},S_{(0)} )^a + V^a S_{(0)} = 0.
\nfr{classy}
For a theory
in which the $Sp(2)$ algebra closes off-shell
the classical solution $S_{(0)}$ takes the form
$$
S_{(0)} = S_0 + \Phi^\star_{Aa} s^a\Phi^A + \half
\Phi^{\star\star}_A \epsilon_{ab} s^as^b \Phi^A
          + F_{AB} \epsilon_{ab} s^a \Phi^A s^b\Phi^B .
\nfr{solzero}
For more complicated
theories like superparticles or superstrings,
$S_{(0)}$ has terms of higher order in the fields
$\Phi^\star_A$, $\Phi^{\star\star}_A$ to compensate those terms
which makes the $Sp(2)$ algebra to close on-shell.
The classical solution to the generating equation
is $Sp(2)$ BRST invariant under modified BRST generators $\tilde s^a$
which satisfy $\tilde s^a \tilde s^b + \tilde s^b \tilde s^a = 0$.

\chapter{Orthosymplectic Structure of the Original BSC Superparticle.}

The original BSC superparticle $S_{BSC}$ and further models are
known to yield the same spectrum as that of $D=10$, $N=1$ super-Yang-Mills
theory [\Ref{elself}].
It is used here as an illustrative example to construct its $Sp(2)$
covariant lagrangian, since the model has an infinitely reducible
algebra. The BSC superparticle action is given by [\Ref{bscone}]
$$
S_0 = \int d\tau [ \pmu\dot x^\mu - i \theta{\psh}\dot\theta
                   - \half e p^2 ].
\nfr{bscparticle}
This
action describes a particle with world-line parametrized by $\tau$
moving through a ten-dimensional $N=1$ superspace with coordinates
$(\xmu ,\theta_A )$. The superparticle action $S_{BSC}$ is invariant
under a $10$ dimensional super-Poincar\'e symmetry
$$
\delta\theta =\epsilon ,\qquad \delta\xmu = i\epsilon\Gamma^\mu\theta ,
\nfr{poincare}
together
with world-line reparametrisations and a local fermionic symmetry
$$\eqalign{\delta\theta =& \psh\kappa ,\qquad
\delta e = 4i\kappa\dot\theta +\dot\xi ,\cr
&\delta\xmu = i\theta\Gamma^\mu\psh\kappa + \xi\pmu .
\cr}\nfr{fermionic}
The Grassmann spinor $\kappa_A$ parametrizes
the local symmetry while $\xi$ parametrizes a linear combination of
world-line
diffeomorphisms and a local {\it trivial} local symmetry.
To construct a covariant $Sp(2)$ orthosymplectic structure for this model,
it is required the formalism of the previuos section since the
classical infinitely reducible gauge algebra $\cal A$
closes on-shell.
It is then defined a superspace $M_S$ to include the classical fields
$\Phi^A = (\xmu ,\pmu ,e,\theta_A )$ and $Sp(2)$ doublets of
anti-fields $\Phi^\star_A$, $\Phi^{\star\star}_A$.
The classical approximation $S_{(0)}$ which satisfies \classy\ is
given by
$$
S_{(0)} = S_{BSC} + S_1 + S_2 + S_3 ,\nfr{orthobsc}
where
$S_{BSC}$ is the classical action of the original superparticle and
$S_1$, $S_2$, and $S_3$ are
$$
\eqalign{
S_1 =\int d\tau [&\theta^\star_a\psh\kappa^a_1 + e^\star_a
(4i\kappa^a_1\dot\theta +\dot c^a )
+ \kappa^\star_{nab} ( (-)^n\psh )(f^{ab}_{\ c}\kappa^c_{n+1}
+\epsilon^{ab}\pi_n ) \cr
&+ x^\star_{\mu a} (i\theta\gamma^\mu\psh\kappa^a_1 + \pmu c)
+ c^\star_{da} (-2i f^{ad}_{rs} \kappa^r_1\psh\kappa^s_1
+\epsilon^{ad}\pi )], \cr}
\nfr{sone}
$$
\eqalign{
S_2 =\int d\tau [&\theta^{\star\star}(-p^2\epsilon_{ab}f^{ab}_{\ c}\kappa^c_2 )
+ e^{\star\star} (-4i\epsilon_{ab}f^{ab}_{\ c}\kappa^c_2\psh\dot\theta
+ 2i\epsilon_{ab}\kappa^a_1\dot{\psh}\kappa^b_1 ) \cr
&+ \kappa^{\star\star}_{n r} ( -p^2 )(\epsilon_{ab}f^{br}_{\ c}
f^{ac}_{\ s}\kappa^s_{n+2} + f^{br}_{\ b}\pi_{n+1}  )  \cr
&+ x^{\star\star}_{\mu} (ip^2\epsilon_{ab} )
( f^{ab}_{\ c}\theta\gammu\kappa^c_2 - \kappa^a_1\gammu\kappa^b_1 )  \cr
&+ c^{\star\star}_e (-4ip^2 )(\epsilon_{ab} f^{be}_{rs}f^{as}_{\ c}
\kappa^r_1\kappa^c_2 + f^{be}_{rb} \kappa^r_1\pi_1 )],
\cr}\nfr{stwo}
and
$$
\eqalign{
S_3 =\int d\tau &\half e^\star_a
[\theta^{\star}_b( -\kappa^c_2 )( 2f^{ab}_{\ c} + \epsilon^{ab}
\epsilon_{rs} f^{rs}_{\ c} ) \cr
&+ x^\star_{\mu b} ( 2i\kappa^a_1\gammu\kappa^b_1
+ i\epsilon^{ab}\epsilon_{rs}\kappa^r_1\gammu\kappa^s_1
- i\theta\gammu\kappa^c_2 ( 2 f^{ab}_{\ c} + \epsilon^{ab}\epsilon_{rs}
f^{rs}_{\ c} ) ) \cr
&+ \kappa^{\star}_{nAb} ( -\kappa^c_{n+2} (\epsilon^{ab}\epsilon_{ps}
f^{sA}_{\ q}f^{pq}_{\ c} + 2 f^{Ab}_{\ s}f^{as}_{\ c} )
-\pi_{n+1} ( 2\epsilon^{as} f^{Ab}_{\ s} + \epsilon^{ab} f^{pA}_{\ p} ) )\cr
&- 4ic^{\star}_{Ab} ( \kappa^r_1\kappa^c_2
(\epsilon^{ab}\epsilon_{ps} f^{sA}_{rq}f^{pq}_{\ c}
+ 2f^{bA}_{rs} f^{as}_{\ c} ) + \kappa^r_1 \pi_1
(\epsilon^{ab} f^{bA}_{rs} + \epsilon^{ab} f^{pA}_{qp} ) )].
\cr}\nfr{sthree}

\sjump

{\it acknowledgments}: I would like to thank S.P. Sorella and C.
Arag{\~ a}o de Carvalho for useful discussions and comments.
I am also grateful to Prof. J.J. Giambiagi for estimulating conversations.}

\bjump\vfill
\eject

\centerline{\capsone REFERENCES}

\sjump
\beginref

\Rref{bon1}{L. Bonora, P. Pasti and M. Tonin, Nuo. Cim. {\bf 63A} (1981) 353.}
\Rref{bon2}{L. Bonora, P. Pasti and M. Tonin, J. Math. Phys. {\bf 23} (1982)
839.}
\Rref{hoyos}{J. Hoyos, M. Quiros, J. Ramirez-Mitterlbrunn and F.J. de Urries,
Nucl. Phys. {\bf B218} (1983) 159.}
\Rref{rogers}{A. Rogers, J. Math. Phys. {\bf 21} (1980) 1352.}
\Rref{baul}{L. Baulieu, Phys. Rep. {\bf 129} (1985) 1.}
\Rref{our}{C. M. Hull, B. Spence and J.L. Vazquez-Bello, Nucl Phys. {\bf B348}
(1991) 108.}
\Rref{espies}{L. Baulieu, W.Siegel and B. Zwiebach, Nucl. Phys. {\bf B287}
(1987) 93; W.Siegel and B. Zwiebach, Nucl. Phys. {\bf B288} (1987) 332;
W. Siegel, {\it Introduction to String Field Theory}, World Scientific (1989).}
\Rref{tieg}{J. Thierry-Mieg, Nucl. Phys. {\bf B261} (1985) 55.}
\Rref{mieg}{J. Thierry-Mieg, J. Math. Phys. {\bf 21} (1980) 2834;
         Nuov. Cim. {\bf 56A} (1980) 396; L. Bonora and M. Tonin, Phys.
         Lett. {\bf 98B} (1981) 48.
         J. Thierry-Mieg and Y. Ne'eman, Ann. Phys. {\bf 123} (1980) 247;
         M. Quiros, F. J. de Urries, J. Hoyos, M. J. Mazon and E.
         Rodr\'{\i}guez,
         J. Math. Phys. {\bf 22} (1981) 1767; L. Bonora, P. Pasti and
         M. Tonin, Nuovo Cim. {\bf A64} (1981) 307;
         R. Delbourgo and P. D. Jarvis,
         J. Phys. A. (Math. Gen.) {\bf 15} (1982) 611; A. Hirshfield and H.
         Leschke, Phys. Lett. {\bf 101B} (1981) 48;
         L. Baulieu, Nucl. Phys. {\bf B241} (1984) 557.}
\Rref{ore}{F.R. Ore and P. Van Niewenhuizen, Nucl. Phys.
{\bf B204} (1982) 317.}
\Rref{gaume}{L. Alvarez-Gaume and L. Baulieu, Nucl. Phys.
{\bf B212} (1983) 255.}
\Rref{spirid}{V.P. Spiridonov, Nucl. Phys. {\bf B308} (1988) 527.}
\Rref{lavrovx}{P. M. Lavrov, Mod. Phys. Lett. {\bf A22} (1991) 2051.}
\Rref{lavrov}{I. A. Batalin, P. M. Lavrov and I. V. Tyutin,
 J. Math. Phys. {\bf 31} (1990) 1487; J. Math. Phys. {\bf 31} (1990) 6;
 J. Math. Phys. {\bf 32} (1991) 532; J. Math. Phys. {\bf 32} (1991) 2513.}
\Rref{hull}{C.M. Hull, Mod. Phys. Lett. {\bf A5} (1990) 1871.}
\Rref{galgebra}{A. Schwarz, Commun. Math. Phys {\bf 158} (1993) 373; Commun.
Math. Phys {\bf 155} (1993) 249; B.H. Lian and G.J. Zuckerman, Commun.
Math. Phys {\bf 154} (1993) 613; M. Gerstenhaber, Ann. Math. {\bf 2}
(1963) 267; M. Penkava and A. Schwarz, Preprint hepth-9212072
.}\Rref{elself}{C.M. Hull and J.L. Vazquez-Bello, Nucl Phys. {\bf B416} (1994)
173; M.B. Green and C.M. Hull, Phys. Lett. {\bf B229} (1989) 215.}
\Rref{bscone}{R.Casalbuoni, Phys. Lett. {\bf B293} (1976) 49; Nuov. Cimm
{\bf 33A} (1976) 389; L. Brink and J.H. Schwarz, Phys. Lett. {\bf B100}
(1981) 310.}

\endref
\ciao